\documentclass[aps,prl,twocolumn,superscriptaddress, floatfix, final]{revtex4-1}
\usepackage{graphicx,amsmath,esint,amssymb,amsfonts,latexsym,color,dcolumn,bm,natbib}
\usepackage[caption=false]{subfig}
\usepackage{verbatim}
\usepackage[%
  pdftex,
  breaklinks=true,
  colorlinks=true,
  linkcolor=blue,
  urlcolor=blue,
  citecolor=blue
]{hyperref}

\def\*#1{\mathbf{#1}}

\newcommand{\Ch}[1]{\color{black}#1\color{black}}

\begin{document}

\title{Intrinsic photon loss at the interface of superconducting devices}

\author{Igor Diniz}  \altaffiliation{igordiniz@ufrrj.br}
\affiliation{Department of Physics and Astronomy, University of Victoria, \\
Victoria, British Columbia, Canada V8W 2Y2}
\affiliation{Centre for Advanced Materials and Related Technology, University of Victoria, Victoria, British Columbia V8W 2Y2, Canada}
\affiliation{Instituto de Ci\^{e}ncias Exatas, Universidade Federal Rural do Rio de Janeiro, Serop\'{e}dica CEP 23890-000, Brazil}

\author{Rog\'erio de Sousa}
\affiliation{Department of Physics and Astronomy, University of Victoria, \\
Victoria, British Columbia, Canada V8W 2Y2}
\affiliation{Centre for Advanced Materials and Related Technology, University of Victoria, Victoria, British Columbia V8W 2Y2, Canada}

\date{\today}

\begin{abstract}
We present a quantum theory of dielectric energy loss arising from the piezoelectric coupling between photons and phonons in superconducting devices. Photon loss is shown to occur predominantly at the interface, where the piezoelectric effect is non-zero even when the materials are perfectly crystalline (epitaxial) and free of two-level system defects. We present explicit numerical calculations for the value of the intrinsic loss tangent at several interfaces to conclude that the $T_1$ of superconducting qubits may reach \Ch{over}~$10^{4}$~$\mu$s if the device is made with defect-free interfaces. 
\end{abstract}
\maketitle

Qubits based on Josephson junctions have come a long way and became one of the most promising devices for quantum information processing. Although coherence times have improved by several orders of magnitude in the past two decades \cite{Devoret2013}, relatively short coherence is still arguably the main obstacle in the implementation of large scale quantum computation. 

Coherence times in state-of-the-art Josephson devices are limited by energy relaxation ($T_1$ decay) \cite{Rigetti2012, Nguyen2019, Place2020} and there are many noise/relaxation sources that can play a role. Understanding the physical origin of these sources is key to make further progress on coherence times. While any excitation with electric dipole moment can contribute to electric (photon) energy loss, a large number of experiments with superconducting resonators provide evidence of loss dominated by extrinsic sources that can be modelled as a bath of two-level-systems (TLSs) \cite{Martinis2005, Shalibo2010, Grabovskij2012, Khalil2014, Skacel2015, Lisenfeld2016, Sarabi2016, Romanenko2017}. The evidence for TLSs is based on the observation that the loss tangent (proportional to the inverse quality factor of the circuit, $1/Q$) always decreases with increasing microwave power, and this can only be explained by TLS saturation. At high power, when TLSs are saturated, the origin of the residual loss is not understood \cite{Skacel2015}. 

An additional mechanism of loss is phonon radiation due to the piezoelectric effect \cite{OConnell2010}. 
It is well known that Josephson junctions radiate phonons at the Josephson frequency, but it is still not clear whether this occurs due to presence of TLSs or due to the piezoelectric effect \cite{Berberich1982}. Ioffe {\it et al.} \cite{Ioffe2004} proposed a mechanism of 
phonon radiation due to piezoelectricity in disordered junctions; assuming the qubit electrical energy was mostly concentrated at the Josephson junction \Ch{and a rough estimate using bulk material parameters}~led to the conclusion that this effect could be responsible for the typical $T_1$ observed in superconducting qubits \cite{Ioffe2004}. 
However, systematic studies of qubit relaxation for varying qubit geometries showed that $1/T_1$ was proportional to the 
electrical energy at the interfaces away from the Josephson junction \cite{Wang2015a, Dial2016}. 
In spite of its ubiquity, the contribution of piezoelectricity to the quality factor of superconducting qubits is not known.

In this Letter we describe a quantum theory of photons and phonons coupled by the piezoelectric effect. In the microwave range the resulting loss is ineffective in large bulk piezoelectric materials. However, the loss is found to be greatly enhanced at piezoelectric substrates with finite thickness, as well as surfaces and interfaces of non-piezoelectric materials, where material discontinuity generally leads to piezoelectricity. 
As a result, dielectric loss due to phonon radiation is an \emph{intrinsic effect}, that is present even when the surfaces, interfaces, and substrates are perfect crystals.

\emph{Quantum theory of photons and phonons coupled by the piezoelectric effect.--} When a photon travels inside an insulator it inevitably has finite lifetime, in that the pure photon is no longer an eigenstate of the material's Hamiltonian.  This occurs because the material has excitations and defects with electric dipole moment. 
The coupling is most effective when the frequency of the photon is resonant with the frequencies of the excitations contributing to the material's polarization $\bm{P}$ (electric dipole moment per volume). In the microwave range a large density of acoustic phonons always satisfies these conditions; the phonons acquire electric dipole moment whenever the material or device lacks inversion symmetry, e.g. due to the presence of an interface or disorder. 

\Ch{As a starting point, we take a single photon mode as the representative for electrical energy stored in a quantum device. Later we generalize to many modes and arbitrary electrical energy distribution. The Hamiltonian for photon plus
phonons is given by}
\begin{equation}
{\cal H}_0=\hbar\Omega\left(a^{\dag}a+\frac{1}{2}\right) + \sum_{\bm{k}}\hbar\omega_k \left(b^{\dag}_{\bm{k}}b_{\bm{k}}+\frac{1}{2}\right), \label{H0}
\end{equation}
where the operator $a^{\dag}$ creates a photon with frequency $\Omega$, and the operator $b^{\dag}_{\bm{k}}$  creates an acoustic phonon with 
wavevector $\bm{k}$ and frequency $\omega_{k}=v |\bm{k}|$, with $v$ the phonon velocity. 
\Ch{The photon electric field operator}
\begin{equation}
\bm{E}=\sqrt{\frac{\hbar\Omega}{2\epsilon V_a}}\left[\bm{\psi}(\bm{r})a+\bm{\psi}^{*}(\bm{r})a^{\dag}\right]
\end{equation}
\Ch{is written in terms of the photon shape vector 
$\bm{\psi}(\bm{r})$}~
which is normalized to the \Ch{photon}~volume, $\int d^3r |\bm{\psi}|^2=V_a$. The constant $\epsilon$ is the microwave frequency dielectric constant, which arises from non-resonant mechanisms such as electronic and optical phonon excitations. 

\Ch{Due to piezoelectricity the phonon electric polarization $\bm{P}$ becomes approximately proportional to the divergence of the phonon displacement operator}, $\bm{P}=\bm{g}\nabla\cdot \bm{u}$ \cite{Snoke2020}. The constant of proportionality $\bm{g}(\bm{r})$
is denoted ``piezoelectric vector''; \Ch{as shown in \cite{Supplemental} this is a function of the coefficients of the piezoelectric tensor}. Here
$\bm{g}(\bm{r})$ is assumed to depend on position in order to describe inhomogeneous systems such as interfaces and junctions. Inserting the usual expression for phonon displacement $\bm{u}$ we get
\begin{equation}
\bm{P}(\bm{r})=\bm{g}(\bm{r}) i\sum_{\bm{k}}\sqrt{\frac{\hbar\omega_{k}}{2\rho V v^2}} \left(
b_{\bm{k}}\textrm{e}^{i\bm{k}\cdot \bm{r}}-
b_{\bm{k}}^{\dag}\textrm{e}^{-i\bm{k}\cdot \bm{r}}
\right),
\end{equation}
where $V$ is the volume of the insulator (e.g. the dielectric substrate, which is assumed to be different than $V_a$, the volume of the photon mode), and $\rho$ is its mass density. 
Note that $\bm{g}$ has the same dimensions as $\bm{P}$, charge/area, and by symmetry it points perpendicular to an interface.

The interaction between photons and phonons is given by
\begin{eqnarray}
{\cal H}_{{\rm int}}&=& -\int d^3 r \bm{P}(\bm{r})\cdot\bm{E}(\bm{r})\nonumber\\
&&=\sum_{\bm{k}} \left(\xi_{\bm{k}}\,a b^{\dag}_{\bm{k}} + {\rm H.c.}\right),
\label{Hint}
\end{eqnarray}
with coupling amplitude
\begin{equation}
\xi_{\bm{k}}=i\sqrt{\frac{\hbar^2 \Omega\omega_{k}}{4\rho \epsilon v^2 V_a V}}\int d^3 r \;\bm{g}(\bm{r})\cdot \bm{\psi}(\bm{r})\textrm{e}^{-i\bm{k}\cdot\bm{r}}.
\label{xik}
\end{equation}
In Eq.~(\ref{Hint}) we neglected terms such as $a b_{\bm{k}}$ and $a^{\dag}b^{\dag}_{\bm{k}}$, because they can't conserve energy so they don't contribute to energy loss. 
The terms that conserve total energy lead to energy dissipation for the photon system, with rate given by
\begin{equation}
{\cal R}_{{\rm diss}}=\hbar\Omega \left( \Gamma_{a~\rightarrow~b} - \Gamma_{b~\rightarrow~a}\right), 
\label{rdiss}
\end{equation}
where $\Gamma_{a~\rightarrow~b}$  is the rate for processes
that convert a photon into a phonon (energy loss), with $\Gamma_{b~\rightarrow~a}$ the opposite process of energy gain. The former and the latter are induced by the terms
$a b^{\dag}_{\bm{k}}$ and $b_{\bm{k}} a^{\dag}$ in Eq.~(\ref{Hint}), respectively. 
Using Fermi's golden rule we get 
\begin{equation}
\Gamma_{a~\rightarrow~b}=\frac{2\pi}{\hbar}\sum_{\bm{k}}\left|\xi_{\bm{k}}\right|^{2}n_a\left(n_k+1\right)\delta\left(\hbar\Omega-\hbar\omega_k\right), 
\label{gamma_ab}
\end{equation}
where $n_a$ and $n_k$ are the number of photons in mode $a$ and the number of phonons in mode $\bm{k}$, respectively.
The expression for $\Gamma_{b~\rightarrow~a}$ is obtained by replacing $n_a(n_k+1)$ for $(n_a+1)n_k$. 

Plugging the amplitudes (\ref{xik}) into Eqs.~(\ref{rdiss})~and~(\ref{gamma_ab}) leads to a general expression for the inverse quality factor $1/Q$, which is the fractional energy lost per cycle:
\begin{eqnarray}
    \frac{1}{Q}&=&\frac{1}{\Omega}\frac{{\cal R}_{{\rm diss}}}{\hbar\Omega\left(n_{a}+\frac{1}{2}\right)}\nonumber\\
    &=& \frac{\Omega^3\left[n_a-n_B(\Omega)\right]}{4\pi\rho v^5\epsilon V_a\left(n_a+\frac{1}{2}\right)}\int d^3 r \int d^3 r'\left[\bm{g}(\bm{r})\cdot\bm{\psi}(\bm{r})\right]\nonumber\\
&&\times{\rm sinc}\left(\frac{\Omega}{v}\left|\bm{r}-\bm{r}'\right|\right)
\left[\bm{g}(\bm{r}')\cdot\bm{\psi}^{*}(\bm{r}')\right],\label{tandeltagen}
\end{eqnarray}
where we assumed the phonons are at thermal equilibrium at some temperature $T$, i.e. their occupation is equal to the Bose distribution $n_B(\Omega)=1/[\exp{(\hbar\Omega/k_BT)}-1]$. If in addition the photon system is also at thermal equilibrium, $n_a$  will also be equal to $n_B(\Omega)$ and Eq.~(\ref{tandeltagen}) will become exactly equal to zero. This shows that Eq.~(\ref{tandeltagen}) satisfies detailed balance. 

It is straightforward to generalize Eq.~(\ref{tandeltagen}) to an arbitrary number of photon modes. The final answer is to replace $\bm{\psi}(\bm{r})$ by 
$\bm{E}(\bm{r})/\sqrt{2\int d^3 r |\bm{E}(\bm{r})|^2/V_a}$, where $\bm{E}(\bm{r})$ is the space-dependent electric field (a classical field). 

\emph{Key role of photon confinement.--} Consider a bulk piezoelectric material so that  $V_a\rightarrow \infty$ and assume that $\bm{g}(\bm{r})$ points along some direction in the crystal with $|\bm{g}(\bm{r})|=g_B$ constant. In this case the photons can be regarded as plane waves, $\bm{\psi}(\bm{r})=\textrm{e}^{i\bm{q}\cdot \bm{r}}\bm{\hat{e}}$, and Eq.~(\ref{xik}) is non-zero only for phonons with $\bm{k}=\bm{q}$ (conservation of momentum). Since Eq.~(\ref{gamma_ab}) requires conservation of energy ($\Omega=\omega_{\bm{k}}$ or $c|\bm{q}|=v|\bm{k}|$), it yields $1/Q=0$ for $\Omega>0$. Therefore, the piezoelectric mechanism yields zero dissipation in bulk. 

\begin{figure*}[t]
\begin{center}
\subfloat[]{\includegraphics[width=0.49\textwidth]{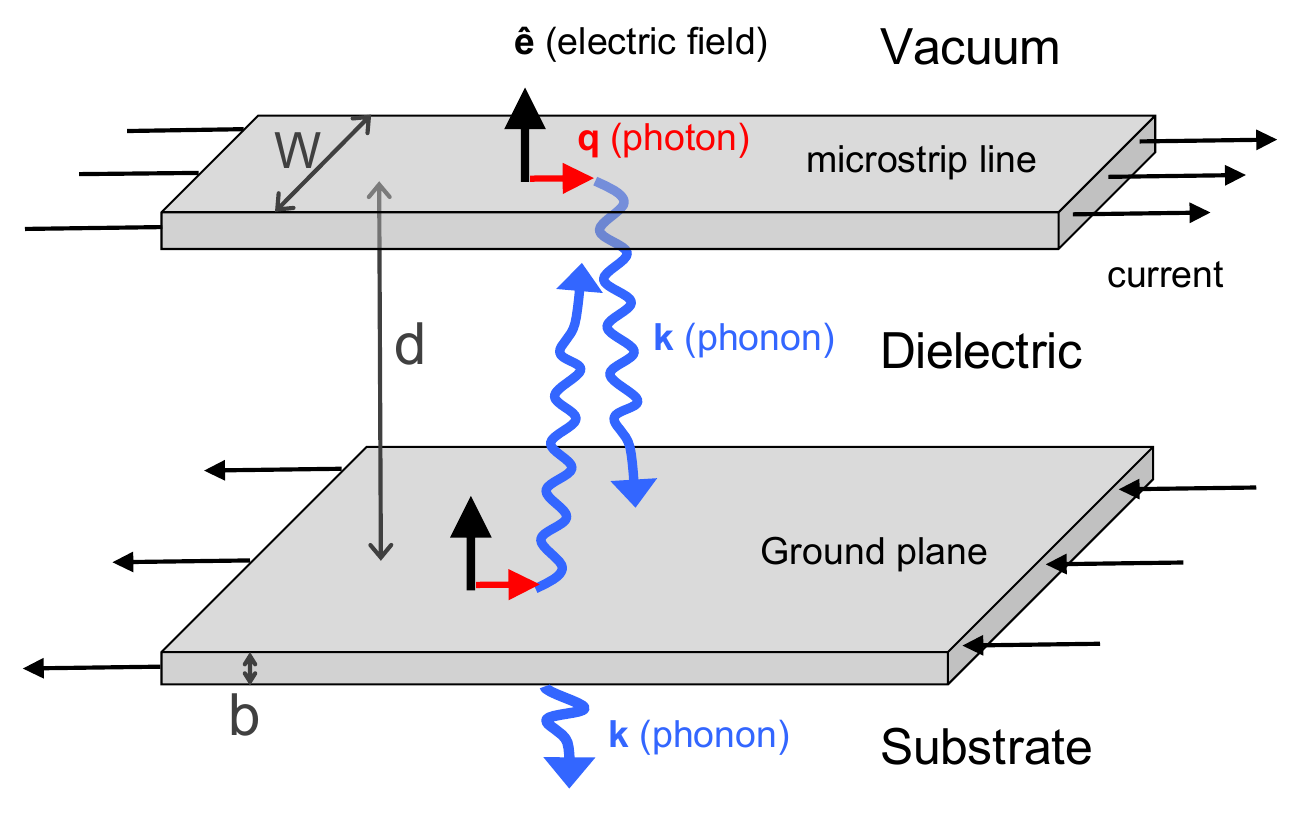}
\label{FigPhononInterference}}
\quad
\subfloat[]{\includegraphics[width=0.45\textwidth]{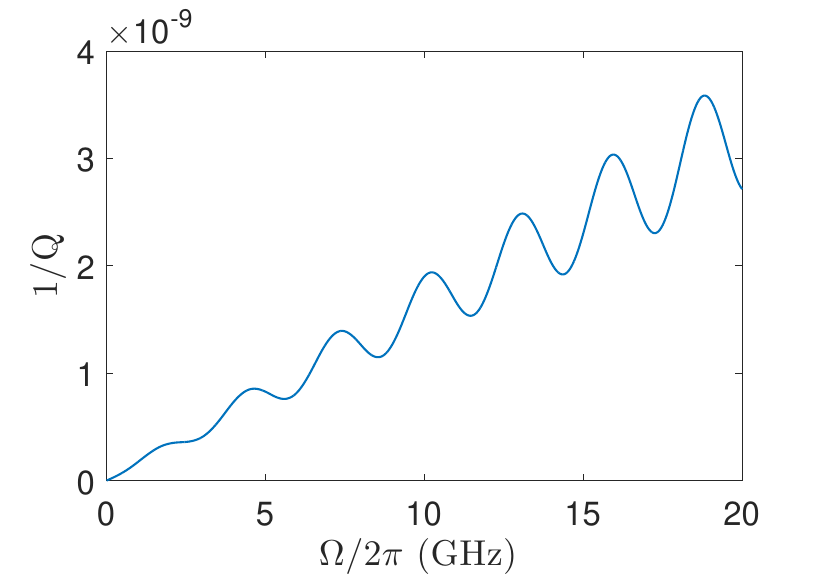}
\label{FigOscillations}}
\end{center}
\caption{(Color online) (a)  Photon loss due to the piezoelectric effect at the interface. A photon travelling in the dielectric waveguide formed by a superconducting microstrip line and ground plane spontaneously decays into an 
acoustic phonon. The selection rules for photon to phonon conversion ensure the phonon propagates nearly perpendicular to the interface, with emission from the top and bottom interfaces interfering with each other. 
(b) Calculated $1/Q$ as a function of photon frequency $\Omega$ for the microstrip line shown in (a). The metal is Al ($0.2$~$\mu$m thick) and the dielectric and substrate is sapphire
(Al$_2$O$_3$), with $W=20$~$\mu$m and $d=2$~$\mu$m. We used the participation ratios calculated in \cite{Wenner2011}, $f_{MV}= 6.5\times 10^{-6}$, $f_{DV}=2.9\times 10^{-4}$, and $f_{DM}=2.9\times 10^{-3}$. The loss is oscillatory as a function of frequency, making evident the presence of phonon interference. These oscillations can be used to distinguish the piezoelectric effect from other sources of loss.\label{fig1}}
\end{figure*}

Now consider what happens in a piezoelectric substrate with large area $A\rightarrow \infty$, but finite thickness $L$. In this case we have $\bm{g}(\bm{r})=g_B \theta(-z)\theta(z+L)\hat{\bm{z}}$, where $\theta(z)$ is the Heaviside step function and $\bm{\hat{z}}$ is the unit vector perpendicular to plane $A$. The photon shape function is assumed to be $\bm{\psi}(\bm{r})=\textrm{e}^{i\bm{q}_{\perp}\cdot \bm{r}}\bm{\hat{z}}$, with photon propagation wavevector $\bm{q}_{\perp}$ perpendicular to $\bm{\hat{z}}$.
Now the phonon-photon momentum conservation in Eq.~(\ref{xik}) is reduced to $\bm{k}_{\perp}=\bm{q}_{\perp}$, with $\bm{k}\cdot \bm{\hat{z}}$ arbitrary.
This freedom allows satisfaction of energy conservation with  $\bm{k}\cdot\bm{\hat{z}}=\pm \Omega \sqrt{1/v^2-1/c^2}\approx \pm\Omega/v$, where $+$ ($-$) denotes a phonon emitted 
along $+\bm{\hat{z}}$ ($-\bm{\hat{z}}$). These considerations allow exact evaluation of Eq.~(\ref{tandeltagen}), leading to 
$1/Q=(L A/V_a) \tan{(\delta_S)}$. 
The prefactor $f_S=(L A/V_a)\leq 1$ is the fraction of total electrical energy at the substrate, denoted participation ratio  \cite{Wang2015a}, and the intrinsic loss tangent for the substrate is given by
\begin{equation}
\tan{(\delta_S)}= \frac{g_{B}^{2}\left[n_a-n_B(\Omega)\right]
}{2 \epsilon \rho v \left(n_a+\frac{1}{2}\right)\Omega L}\sin^{2}{\left(\frac{\Omega L}{v}\right)}.
\label{tandeltaS}
\end{equation}
Note how this is proportional to $1/(\Omega L)$, so it goes to zero in the bulk limit: When either $L\rightarrow \infty$ or $\Omega$ is large enough so that the phonon wavelength is much smaller than $L$, $\lambda_{{\rm phonon}}=2\pi v/\Omega  \ll L$. Moreover, Eq.~(\ref{tandeltaS}) is oscillatory as a function of $\Omega$ and $L$. This is a consequence of phonon interference. This interference, being unique to the piezoelectric mechanism, offers a way to distinguish it from other sources such as extrinsic loss due to TLSs. However, the interference averages out when $|\bm{\psi}(\bm{r})|^2$ varies on the scale of  $\lambda_{\rm{phonon}}$ ($\sim 1$~$\mu$m for $\Omega\sim $~GHz). In the case of spatial variations the sine squared in Eq.~(\ref{tandeltaS}) averages to $1/2$ and we denote the loss tangent by $\langle \tan{(\delta_S)}\rangle$.  The washing out of the interference can be avoided in other geometries such as the stripline which we explore to give a clear signature of the intrinsic piezoelectricity, see Fig.~\ref{fig1}(a).


\begin{table}[ht!]
  \begin{center}
    \caption{Table of interface thickness $t_I$ and piezoelectric coefficients $g_I$ appearing in $\bm{g}(\bm{r})=g_I t_I \delta(z)\bm{\hat{z}}$ for non-piezoelectric materials Al, Nb, and sapphire. Also shown are the parameters for bulk piezoelectric substrates: Coefficients $g_B$ appearing in $|\bm{g}(\bm{r})|=g_B$
    and the assumed substrate thickness $L$ for calculations shown in Table~\ref{table_tandeltaI}.}
    \label{table_gI}
\vspace{2ex}
    \begin{tabular}{l c c c}
\hline\hline
{\bf Metal/Vacuum} &  $t_I$~(\AA) & $g_I$~(C/m$^2$) & Reference\\
\hline
Al & $2.03$ & $0.73$ & Calculated in \cite{Supplemental}\\
Nb & $1.65$ & $0.18$ & Calculated in \cite{Supplemental}\\
\hline
{\bf Dielectric/Vacuum} & & &\\
\hline
Al$_2$O$_3$ & $2.17$   & $0.16$ & First-principles \cite{Georgescu2019}\\
\hline
{\bf Dielectric/Metal} & & & \\
\hline
Al$_2$O$_3$/Al & $2.17$ & $0.06$ & Calculated in \cite{Supplemental}\\
\hline\hline
{\bf Substrate} & $L$ (\AA) & $g_B$ (C/m$^2$) & \\
\hline
SiO$_2$ & $10^3$ & $0.09$ & Measured in \cite{Tarumi2007}\\
Nb$_2$O$_5$ & $10^2$ & $1$ & Estimated\\
\hline\hline
    \end{tabular}
  \end{center}
\end{table}

%

Now consider what happens at the surface or interface of non-piezoelectric materials (with $g_B=0$). The photon electric field induces a screening areal charge density, which changes the effective charge of interface atoms. As a result the materials are subject to extra electric stress within a length scale $t_I$, the surface/interface thickness. A simple calculation \cite{Supplemental} yields $\bm{g}(\bm{r})=g_I t_I \delta(z)\bm{\hat{z}}$, with 
$\bm{\hat{z}}$ the unit vector perpendicular to the interface (pointing from material 1 to 2), and 
$g_I$ depending on the type of surface/interface as shown in Table~\ref{table_gI}.

Explicit calculation of Eq.~(\ref{tandeltagen}) for $\bm{g}(\bm{r})=g_I t_I \delta(z)\bm{\hat{z}}$ and $\bm{\psi}(\bm{r})=\textrm{e}^{i\bm{q}_{\perp}\cdot \bm{r}}\bm{\hat{z}}$ leads to $1/Q=(t_I A/V_a) \tan{(\delta_I)}$, where $f_I=(t_I A/V_a)$ and $\tan{(\delta_I)}$ are the participation ratio and intrinsic loss tangent for the interface, 
\begin{equation}
\tan{(\delta_I)}= \frac{t_I\Omega g_{I}^{2}\left[n_a-n_B(\Omega)\right]
}{4 \epsilon \left(n_a+\frac{1}{2}\right)}\sum_{i=1,2}\frac{1}{\rho_i v_{i}^{3}}.
\label{tandeltaI}
\end{equation}
The last factor in Eq.~(\ref{tandeltaI}) contains parameters for the two interface materials ($i=1,2$). This happens because the phonon propagating along $+\bm{\hat{z}}$ ($-\bm{\hat{z}}$)  moves into material 2 (1). 

\begin{table}[ht!]
  \begin{center}
    \caption{Predicted values for the intrinsic loss tangent for \emph{epitaxial} junctions, interfaces, and substrates. $\tan{(\delta_J)}$ is the Josephson junction loss tangent calculated from Eq.~(\ref{tandeltaJ}) using $V_J=2\times 10^{8}$~\AA$^{3}$. $tan{(\delta_I)}$ is for epitaxial surfaces/interfaces of non-piezoelectric materials Al, Nb, and sapphire, using Eq.~(\ref{tandeltaI}) and Table~\ref{table_gI}. $\langle tan{(\delta_S)}\rangle$ is for piezoelectric substrates quartz (SiO$_2$) and niobium pentoxide (Nb$_2$O$_5$),
    using Eq.~(\ref{tandeltaS}) with $\sin^2\rightarrow 1/2$, for substrate thickness $L$ as in Table~\ref{table_gI}. These values should be compared to the \emph{extrinsic} loss tangent due to amorphous TLSs, $\tan{(\delta_{{\rm TLS}})}\sim 10^{-3}$ \cite{Wang2015a}. 
    }
    \label{table_tandeltaI}
\vspace{2ex}
    \begin{tabular}{l c}
\hline\hline
{\bf Junctions} &  $\tan{(\delta_{J})}$ \\
\hline
Al/Al$_2$O$_3$/Al & $1\times 10^{-7}$\\
Nb/Nb$_2$O$_5$/Nb & $4\times 10^{-4}$ \\
\hline
{\bf Metal/Vacuum} &  $\tan{(\delta_{I})}$ \\
\hline
Al & $2\times 10^{-4}$\\
Nb & $5\times 10^{-6}$ \\
\hline
{\bf Dielectric/Vacuum} & \\
\hline
Al$_2$O$_3$ & $1\times 10^{-7}$ \\
\hline
{\bf Dielectric/Metal} &  \\
\hline
Al$_2$O$_3$/Al & $1\times 10^{-7}$\\
\hline
{\bf Substrate} & $\langle\tan{(\delta_S)}\rangle$\\
\hline
SiO$_2$ & $4\times 10^{-4}$ \\
Nb$_2$O$_5$ & $1\times 10^{-3}$\\
\hline\hline
    \end{tabular}
  \end{center}
\end{table}


For a small Josephson junction with lateral size $\ll v/\Omega$ we may approximate $\bm{g}(\bm{r})=g_I V_J \delta(\bm{r})\bm{\hat{z}}$ and 
${\rm sinc}(\Omega |\bm{r}-\bm{r}'|/v)\approx 1$ in Eq.~(\ref{tandeltagen}) leading to 
$1/Q=(V_J/V_a)\tan{(\delta_J)}$, with
\begin{equation}
    \tan{\left(\delta_J\right)} = \frac{\Omega^3 g_{I}^{2}V_J\left[n_a-n_B(\Omega)\right]}{4\pi\rho v^5\epsilon (n_a+\frac{1}{2})},
\label{tandeltaJ}
\end{equation}
where $V_J$ is the volume of the junction. 
This expression contains an additional prefactor of $\left[n_a-n_B(\Omega)\right]/\left[4\pi (n_a+1/2)\right]$ when compared to the result obtained in \cite{Ioffe2004}. 

Table~\ref{table_tandeltaI} shows explicit calculations of Eq.~(\ref{tandeltaI}) for epitaxial junctions, interfaces, and substrates, assuming $\Omega/2\pi=10$~GHz, $T=10$~mK, $n_a=1$, and material parameters described in \cite{Supplemental}. The table shows a factor of $10^{1}-10^{4}$ decrease in loss tangent can be obtained if the extrinsic mechanism due to interface TLSs is suppressed.

For more complex devices with multiple interfaces and junctions such as qubits, one can use $f_i$ to denote the participation ratio
in each region $i$. \Ch{The contribution from dielectric loss to the rate for energy relaxation of a qubit 
with ground $|0\rangle$ and excited state $|1\rangle$ becomes
\begin{equation}
\frac{1}{T_1}
=\frac{2C}{\hbar}\coth{\left(\frac{\hbar\Omega}{2k_BT}\right)}
\left|\left\langle 1\right|\frac{\partial {\cal H}}{\partial Q} \left|0\right\rangle\right|^2
\sum_i f_i \tan{\left(\delta_i\right)},
\label{t1}
\end{equation}
where ${\cal H}$ is the qubit Hamiltonian, $C$ and $Q$ are its capacitance and associated charge operator, and}~
$\tan{(\delta_i)}$ is the loss tangent calculated from Eqs.~(\ref{tandeltaS})--(\ref{tandeltaJ}). \Ch{For weakly nonlinear oscillators such as transmons, Eq.~(\ref{t1}) reduces to the well known expression $\frac{1}{T_1}=\Omega \sum_i f_i \tan{(\delta_i)}$ when $\hbar\Omega\gg 2 k_BT$ \cite{Wang2015a, Supplemental}}.

\emph{Phonon interference as the experimental signature of the intrinsic piezoelectric effect.--}  For devices with interfaces separated by a distance of the order of the phonon wavelength 
$\lambda_{{\rm phonon}}=2\pi v/\Omega$ ($\sim 1$~$\mu$m for $\Omega\sim $~GHz), the phonons emitted by the interface piezoelectric effect will show signatures of interference. Consider the microstrip line shown in Fig.~\ref{FigPhononInterference}; it can be modelled by 
$\bm{g}(\bm{r})$ equal to a sum of delta functions at each interface. Explicit calculations of Eq.~(\ref{tandeltagen}) 
show that the loss tangent becomes oscillatory as a function of frequency for interface separation $d\sim \lambda_{{\rm phonon}}$, see Fig.~\ref{FigOscillations}. The loss tangent is also oscillatory as a function of $d$. These oscillations can be used to distinguish the interface piezoelectric mechanism from other sources such as extrinsic loss due to TLSs. In the presence of other sources, the amplitude of the oscillations will be diminished. The period of the oscillations is approximately $5$~MHz$\times ({\rm mm}/d)$, so they can be detected with quite small frequency detunings in millimeter sized microwave devices \cite{Paik2011, Wang2015a}. 

\emph{Conclusions.--} We presented a theory of photon loss due to the piezoelectric effect. Our main result is Eq.~(\ref{tandeltagen}), the explicit expression for 
the fraction of photon loss per cycle ($1/Q$) in a general inhomogeneous structure. 

We showed that piezoelectric loss does not occur in bulk materials, it only occurs in the presence of photon confinement. This includes piezoelectric substrates with finite thickness (or equivalently finite photon penetration depth),  as well as interfaces and junctions made of materials that are non-piezoelectric in the bulk.

In current devices, photon loss is dominated by the presence of extrinsic TLS defects with localized dipole moment. Substantial effort is underway \cite{Place2020} to make devices with epitaxial interfaces, free of TLS defects. For these perfect devices, piezoelectricity provides the ultimate loss mechanism: Even perfect interfaces made of non-piezoelectric materials become piezoelectric because of inversion symmetry breaking. Table~\ref{table_tandeltaI} shows explicit numerical predictions of the intrinsic loss tangent in several different crystalline junctions, interfaces, and substrates. 

Using typical transmon qubit participation ratios   $f_{MV}=1\times 10^{-5}$, $f_{DV}=1\times 10^{-4}$, $f_{DM}=1\times 10^{-4}$ and $f_{JJ}=2\times 10^{-4}$ (design A from \cite{Wang2015a}) we predict that a device made of epitaxial aluminum and sapphire will attain $1/Q\sim 10^{-9}$ at $\Omega/2\pi=10$~GHz, with most of the loss occurring at the aluminum/vacuum surface. 
Therefore, superconducting qubits with optimal interfaces 
can reach $T_1$ up to $10^4$~$\mu$s, above the threshold for quantum error correction \cite{Devoret2013}. 
\Ch{Even longer $T_1$ can be reached for the fluxonium qubit \cite{Nguyen2019,Lin2018,Earnest2018} whose design minimizes the quantum matrix element appearing in Eq.~(\ref{t1})}. 

\begin{acknowledgments}
We acknowledge financial support from NSERC (Canada) through its Discovery (RGPIN-2015-03938) and Collaborative Research and Development programs (CRDPJ 478366-14). We thank M. H. Amin, A. N. Cleland, T. Lanting, M. Mariantoni, T. Juginger, and T. Tiedje for useful discussions.
\end{acknowledgments}

\bibliography{IntrinsicEnergyLossInterface}

\begin{thebibliography}{30}%
\makeatletter
\providecommand \@ifxundefined [1]{%
 \@ifx{#1\undefined}
}%
\providecommand \@ifnum [1]{%
 \ifnum #1\expandafter \@firstoftwo
 \else \expandafter \@secondoftwo
 \fi
}%
\providecommand \@ifx [1]{%
 \ifx #1\expandafter \@firstoftwo
 \else \expandafter \@secondoftwo
 \fi
}%
\providecommand \natexlab [1]{#1}%
\providecommand \enquote  [1]{``#1''}%
\providecommand \bibnamefont  [1]{#1}%
\providecommand \bibfnamefont [1]{#1}%
\providecommand \citenamefont [1]{#1}%
\providecommand \href@noop [0]{\@secondoftwo}%
\providecommand \href [0]{\begingroup \@sanitize@url \@href}%
\providecommand \@href[1]{\@@startlink{#1}\@@href}%
\providecommand \@@href[1]{\endgroup#1\@@endlink}%
\providecommand \@sanitize@url [0]{\catcode `\\12\catcode `\$12\catcode
  `\&12\catcode `\#12\catcode `\^12\catcode `\_12\catcode `\%12\relax}%
\providecommand \@@startlink[1]{}%
\providecommand \@@endlink[0]{}%
\providecommand \url  [0]{\begingroup\@sanitize@url \@url }%
\providecommand \@url [1]{\endgroup\@href {#1}{\urlprefix }}%
\providecommand \urlprefix  [0]{URL }%
\providecommand \Eprint [0]{\href }%
\providecommand \doibase [0]{http://dx.doi.org/}%
\providecommand \selectlanguage [0]{\@gobble}%
\providecommand \bibinfo  [0]{\@secondoftwo}%
\providecommand \bibfield  [0]{\@secondoftwo}%
\providecommand \translation [1]{[#1]}%
\providecommand \BibitemOpen [0]{}%
\providecommand \bibitemStop [0]{}%
\providecommand \bibitemNoStop [0]{.\EOS\space}%
\providecommand \EOS [0]{\spacefactor3000\relax}%
\providecommand \BibitemShut  [1]{\csname bibitem#1\endcsname}%
\let\auto@bib@innerbib\@empty
\bibitem [{\citenamefont {Devoret}\ and\ \citenamefont
  {Schoelkopf}(2013)}]{Devoret2013}%
  \BibitemOpen
  \bibfield  {author} {\bibinfo {author} {\bibfnamefont {M.~H.}\ \bibnamefont
  {Devoret}}\ and\ \bibinfo {author} {\bibfnamefont {R.~J.}\ \bibnamefont
  {Schoelkopf}},\ }\href {\doibase 10.1126/science.1231930} {\bibfield
  {journal} {\bibinfo  {journal} {Science}\ }\textbf {\bibinfo {volume}
  {339}},\ \bibinfo {pages} {1169} (\bibinfo {year} {2013})}\BibitemShut
  {NoStop}%
\bibitem [{\citenamefont {Rigetti}\ \emph {et~al.}(2012)\citenamefont
  {Rigetti}, \citenamefont {Gambetta}, \citenamefont {Poletto}, \citenamefont
  {Plourde}, \citenamefont {Chow}, \citenamefont {C{\'{o}}rcoles},
  \citenamefont {Smolin}, \citenamefont {Merkel}, \citenamefont {Rozen},
  \citenamefont {Keefe}, \citenamefont {Rothwell}, \citenamefont {Ketchen},\
  and\ \citenamefont {Steffen}}]{Rigetti2012}%
  \BibitemOpen
  \bibfield  {author} {\bibinfo {author} {\bibfnamefont {C.}~\bibnamefont
  {Rigetti}}, \bibinfo {author} {\bibfnamefont {J.~M.}\ \bibnamefont
  {Gambetta}}, \bibinfo {author} {\bibfnamefont {S.}~\bibnamefont {Poletto}},
  \bibinfo {author} {\bibfnamefont {B.~L.~T.}\ \bibnamefont {Plourde}},
  \bibinfo {author} {\bibfnamefont {J.~M.}\ \bibnamefont {Chow}}, \bibinfo
  {author} {\bibfnamefont {A.~D.}\ \bibnamefont {C{\'{o}}rcoles}}, \bibinfo
  {author} {\bibfnamefont {J.~A.}\ \bibnamefont {Smolin}}, \bibinfo {author}
  {\bibfnamefont {S.~T.}\ \bibnamefont {Merkel}}, \bibinfo {author}
  {\bibfnamefont {J.~R.}\ \bibnamefont {Rozen}}, \bibinfo {author}
  {\bibfnamefont {G.~A.}\ \bibnamefont {Keefe}}, \bibinfo {author}
  {\bibfnamefont {M.~B.}\ \bibnamefont {Rothwell}}, \bibinfo {author}
  {\bibfnamefont {M.~B.}\ \bibnamefont {Ketchen}}, \ and\ \bibinfo {author}
  {\bibfnamefont {M.}~\bibnamefont {Steffen}},\ }\href {\doibase
  10.1103/PhysRevB.86.100506} {\bibfield  {journal} {\bibinfo  {journal} {Phys.
  Rev. B}\ }\textbf {\bibinfo {volume} {86}},\ \bibinfo {pages} {100506}
  (\bibinfo {year} {2012})}\BibitemShut {NoStop}%
\bibitem [{\citenamefont {Nguyen}\ \emph {et~al.}(2019)\citenamefont {Nguyen},
  \citenamefont {Lin}, \citenamefont {Somoroff}, \citenamefont {Mencia},
  \citenamefont {Grabon},\ and\ \citenamefont {Manucharyan}}]{Nguyen2019}%
  \BibitemOpen
  \bibfield  {author} {\bibinfo {author} {\bibfnamefont {L.~B.}\ \bibnamefont
  {Nguyen}}, \bibinfo {author} {\bibfnamefont {Y.~H.}\ \bibnamefont {Lin}},
  \bibinfo {author} {\bibfnamefont {A.}~\bibnamefont {Somoroff}}, \bibinfo
  {author} {\bibfnamefont {R.}~\bibnamefont {Mencia}}, \bibinfo {author}
  {\bibfnamefont {N.}~\bibnamefont {Grabon}}, \ and\ \bibinfo {author}
  {\bibfnamefont {V.~E.}\ \bibnamefont {Manucharyan}},\ }\href {\doibase
  10.1103/PhysRevX.9.041041} {\bibfield  {journal} {\bibinfo  {journal} {Phys.
  Rev. X}\ }\textbf {\bibinfo {volume} {9}},\ \bibinfo {pages} {041041}
  (\bibinfo {year} {2019})}\BibitemShut {NoStop}%
\bibitem [{\citenamefont {Place}\ \emph {et~al.}()\citenamefont {Place},
  \citenamefont {Rodgers}, \citenamefont {Mundada}, \citenamefont {Smitham},
  \citenamefont {Fitzpatrick}, \citenamefont {Leng}, \citenamefont {Premkumar},
  \citenamefont {Bryon}, \citenamefont {Sussman}, \citenamefont {Cheng},
  \citenamefont {Madhavan}, \citenamefont {Babla}, \citenamefont {Jaeck},
  \citenamefont {Gyenis}, \citenamefont {Yao}, \citenamefont {Cava},
  \citenamefont {de~Leon},\ and\ \citenamefont {Houck}}]{Place2020}%
  \BibitemOpen
  \bibfield  {author} {\bibinfo {author} {\bibfnamefont {A.~P.~M.}\
  \bibnamefont {Place}}, \bibinfo {author} {\bibfnamefont {L.~V.~H.}\
  \bibnamefont {Rodgers}}, \bibinfo {author} {\bibfnamefont {P.}~\bibnamefont
  {Mundada}}, \bibinfo {author} {\bibfnamefont {B.~M.}\ \bibnamefont
  {Smitham}}, \bibinfo {author} {\bibfnamefont {M.}~\bibnamefont
  {Fitzpatrick}}, \bibinfo {author} {\bibfnamefont {Z.}~\bibnamefont {Leng}},
  \bibinfo {author} {\bibfnamefont {A.}~\bibnamefont {Premkumar}}, \bibinfo
  {author} {\bibfnamefont {J.}~\bibnamefont {Bryon}}, \bibinfo {author}
  {\bibfnamefont {S.}~\bibnamefont {Sussman}}, \bibinfo {author} {\bibfnamefont
  {G.}~\bibnamefont {Cheng}}, \bibinfo {author} {\bibfnamefont
  {T.}~\bibnamefont {Madhavan}}, \bibinfo {author} {\bibfnamefont {H.~K.}\
  \bibnamefont {Babla}}, \bibinfo {author} {\bibfnamefont {B.}~\bibnamefont
  {Jaeck}}, \bibinfo {author} {\bibfnamefont {A.}~\bibnamefont {Gyenis}},
  \bibinfo {author} {\bibfnamefont {N.}~\bibnamefont {Yao}}, \bibinfo {author}
  {\bibfnamefont {R.~J.}\ \bibnamefont {Cava}}, \bibinfo {author}
  {\bibfnamefont {N.~P.}\ \bibnamefont {de~Leon}}, \ and\ \bibinfo {author}
  {\bibfnamefont {A.~A.}\ \bibnamefont {Houck}},\ }\href
  {http://arxiv.org/abs/2003.00024} {\ }\Eprint
  {http://arxiv.org/abs/2003.00024} {arXiv:2003.00024} \BibitemShut {NoStop}%
\bibitem [{\citenamefont {Martinis}\ \emph {et~al.}(2005)\citenamefont
  {Martinis}, \citenamefont {Cooper}, \citenamefont {McDermott}, \citenamefont
  {Steffen}, \citenamefont {Ansmann}, \citenamefont {Osborn}, \citenamefont
  {Cicak}, \citenamefont {Oh}, \citenamefont {Pappas}, \citenamefont
  {Simmonds},\ and\ \citenamefont {Yu}}]{Martinis2005}%
  \BibitemOpen
  \bibfield  {author} {\bibinfo {author} {\bibfnamefont {J.~M.}\ \bibnamefont
  {Martinis}}, \bibinfo {author} {\bibfnamefont {K.~B.}\ \bibnamefont
  {Cooper}}, \bibinfo {author} {\bibfnamefont {R.}~\bibnamefont {McDermott}},
  \bibinfo {author} {\bibfnamefont {M.}~\bibnamefont {Steffen}}, \bibinfo
  {author} {\bibfnamefont {M.}~\bibnamefont {Ansmann}}, \bibinfo {author}
  {\bibfnamefont {K.~D.}\ \bibnamefont {Osborn}}, \bibinfo {author}
  {\bibfnamefont {K.}~\bibnamefont {Cicak}}, \bibinfo {author} {\bibfnamefont
  {S.}~\bibnamefont {Oh}}, \bibinfo {author} {\bibfnamefont {D.~P.}\
  \bibnamefont {Pappas}}, \bibinfo {author} {\bibfnamefont {R.~W.}\
  \bibnamefont {Simmonds}}, \ and\ \bibinfo {author} {\bibfnamefont {C.~C.}\
  \bibnamefont {Yu}},\ }\href {\doibase 10.1103/PhysRevLett.95.210503}
  {\bibfield  {journal} {\bibinfo  {journal} {Phys. Rev. Lett.}\ }\textbf
  {\bibinfo {volume} {95}},\ \bibinfo {pages} {210503} (\bibinfo {year}
  {2005})}\BibitemShut {NoStop}%
\bibitem [{\citenamefont {Shalibo}\ \emph {et~al.}(2010)\citenamefont
  {Shalibo}, \citenamefont {Rofe}, \citenamefont {Shwa}, \citenamefont
  {Zeides}, \citenamefont {Neeley}, \citenamefont {Martinis},\ and\
  \citenamefont {Katz}}]{Shalibo2010}%
  \BibitemOpen
  \bibfield  {author} {\bibinfo {author} {\bibfnamefont {Y.}~\bibnamefont
  {Shalibo}}, \bibinfo {author} {\bibfnamefont {Y.}~\bibnamefont {Rofe}},
  \bibinfo {author} {\bibfnamefont {D.}~\bibnamefont {Shwa}}, \bibinfo {author}
  {\bibfnamefont {F.}~\bibnamefont {Zeides}}, \bibinfo {author} {\bibfnamefont
  {M.}~\bibnamefont {Neeley}}, \bibinfo {author} {\bibfnamefont {J.~M.}\
  \bibnamefont {Martinis}}, \ and\ \bibinfo {author} {\bibfnamefont
  {N.}~\bibnamefont {Katz}},\ }\href {\doibase 10.1103/PhysRevLett.105.177001}
  {\bibfield  {journal} {\bibinfo  {journal} {Phys. Rev. Lett.}\ }\textbf
  {\bibinfo {volume} {105}},\ \bibinfo {pages} {177001} (\bibinfo {year}
  {2010})}\BibitemShut {NoStop}%
\bibitem [{\citenamefont {Grabovskij}\ \emph {et~al.}(2012)\citenamefont
  {Grabovskij}, \citenamefont {Peichl}, \citenamefont {Lisenfeld},
  \citenamefont {Weiss},\ and\ \citenamefont {Ustinov}}]{Grabovskij2012}%
  \BibitemOpen
  \bibfield  {author} {\bibinfo {author} {\bibfnamefont {G.~J.}\ \bibnamefont
  {Grabovskij}}, \bibinfo {author} {\bibfnamefont {T.}~\bibnamefont {Peichl}},
  \bibinfo {author} {\bibfnamefont {J.}~\bibnamefont {Lisenfeld}}, \bibinfo
  {author} {\bibfnamefont {G.}~\bibnamefont {Weiss}}, \ and\ \bibinfo {author}
  {\bibfnamefont {A.~V.}\ \bibnamefont {Ustinov}},\ }\href {\doibase
  10.1126/science.1226487} {\bibfield  {journal} {\bibinfo  {journal}
  {Science}\ }\textbf {\bibinfo {volume} {338}},\ \bibinfo {pages} {232}
  (\bibinfo {year} {2012})}\BibitemShut {NoStop}%
\bibitem [{\citenamefont {Khalil}\ \emph {et~al.}(2014)\citenamefont {Khalil},
  \citenamefont {Gladchenko}, \citenamefont {Stoutimore}, \citenamefont
  {Wellstood}, \citenamefont {Burin},\ and\ \citenamefont
  {Osborn}}]{Khalil2014}%
  \BibitemOpen
  \bibfield  {author} {\bibinfo {author} {\bibfnamefont {M.~S.}\ \bibnamefont
  {Khalil}}, \bibinfo {author} {\bibfnamefont {S.}~\bibnamefont {Gladchenko}},
  \bibinfo {author} {\bibfnamefont {M.~J.}\ \bibnamefont {Stoutimore}},
  \bibinfo {author} {\bibfnamefont {F.~C.}\ \bibnamefont {Wellstood}}, \bibinfo
  {author} {\bibfnamefont {A.~L.}\ \bibnamefont {Burin}}, \ and\ \bibinfo
  {author} {\bibfnamefont {K.~D.}\ \bibnamefont {Osborn}},\ }\href {\doibase
  10.1103/PhysRevB.90.100201} {\bibfield  {journal} {\bibinfo  {journal} {Phys.
  Rev. B}\ }\textbf {\bibinfo {volume} {90}},\ \bibinfo {pages} {100201(R)}
  (\bibinfo {year} {2014})}\BibitemShut {NoStop}%
\bibitem [{\citenamefont {Skacel}\ \emph {et~al.}(2015)\citenamefont {Skacel},
  \citenamefont {Kaiser}, \citenamefont {Wuensch}, \citenamefont {Rotzinger},
  \citenamefont {Lukashenko}, \citenamefont {Jerger}, \citenamefont {Weiss},
  \citenamefont {Siegel},\ and\ \citenamefont {Ustinov}}]{Skacel2015}%
  \BibitemOpen
  \bibfield  {author} {\bibinfo {author} {\bibfnamefont {S.~T.}\ \bibnamefont
  {Skacel}}, \bibinfo {author} {\bibfnamefont {C.}~\bibnamefont {Kaiser}},
  \bibinfo {author} {\bibfnamefont {S.}~\bibnamefont {Wuensch}}, \bibinfo
  {author} {\bibfnamefont {H.}~\bibnamefont {Rotzinger}}, \bibinfo {author}
  {\bibfnamefont {A.}~\bibnamefont {Lukashenko}}, \bibinfo {author}
  {\bibfnamefont {M.}~\bibnamefont {Jerger}}, \bibinfo {author} {\bibfnamefont
  {G.}~\bibnamefont {Weiss}}, \bibinfo {author} {\bibfnamefont
  {M.}~\bibnamefont {Siegel}}, \ and\ \bibinfo {author} {\bibfnamefont {A.~V.}\
  \bibnamefont {Ustinov}},\ }\href {\doibase 10.1063/1.4905149} {\bibfield
  {journal} {\bibinfo  {journal} {Appl. Phys. Lett.}\ }\textbf {\bibinfo
  {volume} {106}},\ \bibinfo {pages} {022603} (\bibinfo {year}
  {2015})}\BibitemShut {NoStop}%
\bibitem [{\citenamefont {Lisenfeld}\ \emph {et~al.}(2016)\citenamefont
  {Lisenfeld}, \citenamefont {Bilmes}, \citenamefont {Matityahu}, \citenamefont
  {Zanker}, \citenamefont {Marthaler}, \citenamefont {Schechter}, \citenamefont
  {Schon}, \citenamefont {Shnirman}, \citenamefont {Weiss},\ and\ \citenamefont
  {Ustinov}}]{Lisenfeld2016}%
  \BibitemOpen
  \bibfield  {author} {\bibinfo {author} {\bibfnamefont {J.}~\bibnamefont
  {Lisenfeld}}, \bibinfo {author} {\bibfnamefont {A.}~\bibnamefont {Bilmes}},
  \bibinfo {author} {\bibfnamefont {S.}~\bibnamefont {Matityahu}}, \bibinfo
  {author} {\bibfnamefont {S.}~\bibnamefont {Zanker}}, \bibinfo {author}
  {\bibfnamefont {M.}~\bibnamefont {Marthaler}}, \bibinfo {author}
  {\bibfnamefont {M.}~\bibnamefont {Schechter}}, \bibinfo {author}
  {\bibfnamefont {G.}~\bibnamefont {Schon}}, \bibinfo {author} {\bibfnamefont
  {A.}~\bibnamefont {Shnirman}}, \bibinfo {author} {\bibfnamefont
  {G.}~\bibnamefont {Weiss}}, \ and\ \bibinfo {author} {\bibfnamefont {A.~V.}\
  \bibnamefont {Ustinov}},\ }\href {\doibase 10.1038/srep23786} {\bibfield
  {journal} {\bibinfo  {journal} {Sci. Rep.}\ }\textbf {\bibinfo {volume}
  {6}},\ \bibinfo {pages} {23786} (\bibinfo {year} {2016})}\BibitemShut
  {NoStop}%
\bibitem [{\citenamefont {Sarabi}\ \emph {et~al.}(2016)\citenamefont {Sarabi},
  \citenamefont {Ramanayaka}, \citenamefont {Burin}, \citenamefont
  {Wellstood},\ and\ \citenamefont {Osborn}}]{Sarabi2016}%
  \BibitemOpen
  \bibfield  {author} {\bibinfo {author} {\bibfnamefont {B.}~\bibnamefont
  {Sarabi}}, \bibinfo {author} {\bibfnamefont {A.~N.}\ \bibnamefont
  {Ramanayaka}}, \bibinfo {author} {\bibfnamefont {A.~L.}\ \bibnamefont
  {Burin}}, \bibinfo {author} {\bibfnamefont {F.~C.}\ \bibnamefont
  {Wellstood}}, \ and\ \bibinfo {author} {\bibfnamefont {K.~D.}\ \bibnamefont
  {Osborn}},\ }\href {\doibase 10.1103/PhysRevLett.116.167002} {\bibfield
  {journal} {\bibinfo  {journal} {Phys. Rev. Lett.}\ }\textbf {\bibinfo
  {volume} {116}},\ \bibinfo {pages} {167002} (\bibinfo {year}
  {2016})}\BibitemShut {NoStop}%
\bibitem [{\citenamefont {Romanenko}\ and\ \citenamefont
  {Schuster}(2017)}]{Romanenko2017}%
  \BibitemOpen
  \bibfield  {author} {\bibinfo {author} {\bibfnamefont {A.}~\bibnamefont
  {Romanenko}}\ and\ \bibinfo {author} {\bibfnamefont {D.~I.}\ \bibnamefont
  {Schuster}},\ }\href {\doibase 10.1103/PhysRevLett.119.264801} {\bibfield
  {journal} {\bibinfo  {journal} {Phys. Rev. Lett.}\ }\textbf {\bibinfo
  {volume} {119}},\ \bibinfo {pages} {264801} (\bibinfo {year}
  {2017})}\BibitemShut {NoStop}%
\bibitem [{\citenamefont {O'Connell}\ \emph {et~al.}(2010)\citenamefont
  {O'Connell}, \citenamefont {Hofheinz}, \citenamefont {Ansmann}, \citenamefont
  {Bialczak}, \citenamefont {Lenander}, \citenamefont {Lucero}, \citenamefont
  {Neeley}, \citenamefont {Sank}, \citenamefont {Wang}, \citenamefont {Weides},
  \citenamefont {Wenner}, \citenamefont {Martinis},\ and\ \citenamefont
  {Cleland}}]{OConnell2010}%
  \BibitemOpen
  \bibfield  {author} {\bibinfo {author} {\bibfnamefont {A.~D.}\ \bibnamefont
  {O'Connell}}, \bibinfo {author} {\bibfnamefont {M.}~\bibnamefont {Hofheinz}},
  \bibinfo {author} {\bibfnamefont {M.}~\bibnamefont {Ansmann}}, \bibinfo
  {author} {\bibfnamefont {R.~C.}\ \bibnamefont {Bialczak}}, \bibinfo {author}
  {\bibfnamefont {M.}~\bibnamefont {Lenander}}, \bibinfo {author}
  {\bibfnamefont {E.}~\bibnamefont {Lucero}}, \bibinfo {author} {\bibfnamefont
  {M.}~\bibnamefont {Neeley}}, \bibinfo {author} {\bibfnamefont
  {D.}~\bibnamefont {Sank}}, \bibinfo {author} {\bibfnamefont {H.}~\bibnamefont
  {Wang}}, \bibinfo {author} {\bibfnamefont {M.}~\bibnamefont {Weides}},
  \bibinfo {author} {\bibfnamefont {J.}~\bibnamefont {Wenner}}, \bibinfo
  {author} {\bibfnamefont {J.~M.}\ \bibnamefont {Martinis}}, \ and\ \bibinfo
  {author} {\bibfnamefont {A.~N.}\ \bibnamefont {Cleland}},\ }\href {\doibase
  10.1038/nature08967} {\bibfield  {journal} {\bibinfo  {journal} {Nature}\
  }\textbf {\bibinfo {volume} {464}},\ \bibinfo {pages} {697} (\bibinfo {year}
  {2010})}\BibitemShut {NoStop}%
\bibitem [{\citenamefont {Berberich}\ \emph {et~al.}(1982)\citenamefont
  {Berberich}, \citenamefont {Buemann},\ and\ \citenamefont
  {Kinder}}]{Berberich1982}%
  \BibitemOpen
  \bibfield  {author} {\bibinfo {author} {\bibfnamefont {P.}~\bibnamefont
  {Berberich}}, \bibinfo {author} {\bibfnamefont {R.}~\bibnamefont {Buemann}},
  \ and\ \bibinfo {author} {\bibfnamefont {H.}~\bibnamefont {Kinder}},\ }\href
  {\doibase 10.1103/PhysRevLett.49.1500} {\bibfield  {journal} {\bibinfo
  {journal} {Phys. Rev. Lett.}\ }\textbf {\bibinfo {volume} {49}},\ \bibinfo
  {pages} {1500} (\bibinfo {year} {1982})}\BibitemShut {NoStop}%
\bibitem [{\citenamefont {Ioffe}\ \emph {et~al.}(2004)\citenamefont {Ioffe},
  \citenamefont {Geshkenbein}, \citenamefont {Helm},\ and\ \citenamefont
  {Blatter}}]{Ioffe2004}%
  \BibitemOpen
  \bibfield  {author} {\bibinfo {author} {\bibfnamefont {L.~B.}\ \bibnamefont
  {Ioffe}}, \bibinfo {author} {\bibfnamefont {V.~B.}\ \bibnamefont
  {Geshkenbein}}, \bibinfo {author} {\bibfnamefont {C.}~\bibnamefont {Helm}}, \
  and\ \bibinfo {author} {\bibfnamefont {G.}~\bibnamefont {Blatter}},\ }\href
  {\doibase 10.1103/PhysRevLett.93.057001} {\bibfield  {journal} {\bibinfo
  {journal} {Phys. Rev. Lett.}\ }\textbf {\bibinfo {volume} {93}},\ \bibinfo
  {pages} {057001} (\bibinfo {year} {2004})}\BibitemShut {NoStop}%
\bibitem [{\citenamefont {Wang}\ \emph {et~al.}(2015)\citenamefont {Wang},
  \citenamefont {Axline}, \citenamefont {Gao}, \citenamefont {Brecht},
  \citenamefont {Chu}, \citenamefont {Frunzio}, \citenamefont {Devoret},\ and\
  \citenamefont {Schoelkopf}}]{Wang2015a}%
  \BibitemOpen
  \bibfield  {author} {\bibinfo {author} {\bibfnamefont {C.}~\bibnamefont
  {Wang}}, \bibinfo {author} {\bibfnamefont {C.}~\bibnamefont {Axline}},
  \bibinfo {author} {\bibfnamefont {Y.~Y.}\ \bibnamefont {Gao}}, \bibinfo
  {author} {\bibfnamefont {T.}~\bibnamefont {Brecht}}, \bibinfo {author}
  {\bibfnamefont {Y.}~\bibnamefont {Chu}}, \bibinfo {author} {\bibfnamefont
  {L.}~\bibnamefont {Frunzio}}, \bibinfo {author} {\bibfnamefont {M.~H.}\
  \bibnamefont {Devoret}}, \ and\ \bibinfo {author} {\bibfnamefont {R.~J.}\
  \bibnamefont {Schoelkopf}},\ }\href {\doibase 10.1063/1.4934486} {\bibfield
  {journal} {\bibinfo  {journal} {Appl. Phys. Lett.}\ }\textbf {\bibinfo
  {volume} {107}},\ \bibinfo {pages} {162601} (\bibinfo {year}
  {2015})}\BibitemShut {NoStop}%
\bibitem [{\citenamefont {Dial}\ \emph {et~al.}(2016)\citenamefont {Dial},
  \citenamefont {McClure}, \citenamefont {Poletto}, \citenamefont {Keefe},
  \citenamefont {Rothwell}, \citenamefont {Gambetta}, \citenamefont {Abraham},
  \citenamefont {Chow},\ and\ \citenamefont {Steffen}}]{Dial2016}%
  \BibitemOpen
  \bibfield  {author} {\bibinfo {author} {\bibfnamefont {O.}~\bibnamefont
  {Dial}}, \bibinfo {author} {\bibfnamefont {D.~T.}\ \bibnamefont {McClure}},
  \bibinfo {author} {\bibfnamefont {S.}~\bibnamefont {Poletto}}, \bibinfo
  {author} {\bibfnamefont {G.~A.}\ \bibnamefont {Keefe}}, \bibinfo {author}
  {\bibfnamefont {M.~B.}\ \bibnamefont {Rothwell}}, \bibinfo {author}
  {\bibfnamefont {J.~M.}\ \bibnamefont {Gambetta}}, \bibinfo {author}
  {\bibfnamefont {D.~W.}\ \bibnamefont {Abraham}}, \bibinfo {author}
  {\bibfnamefont {J.~M.}\ \bibnamefont {Chow}}, \ and\ \bibinfo {author}
  {\bibfnamefont {M.}~\bibnamefont {Steffen}},\ }\href {\doibase
  10.1088/0953-2048/29/4/044001} {\bibfield  {journal} {\bibinfo  {journal}
  {Supercond. Sci. Technol.}\ }\textbf {\bibinfo {volume} {29}},\ \bibinfo
  {pages} {044001} (\bibinfo {year} {2016})}\BibitemShut {NoStop}%
\bibitem [{\citenamefont {Snoke}(2020)}]{Snoke2020}%
  \BibitemOpen
  \bibfield  {author} {\bibinfo {author} {\bibfnamefont {D.~W.}\ \bibnamefont
  {Snoke}},\ }\href@noop {} {\emph {\bibinfo {title} {Solid State Physics:
  Essential Concepts}}},\ \bibinfo {edition} {2nd}\ ed.\ (\bibinfo  {publisher}
  {Cambridge University Press, U.K.},\ \bibinfo {year} {2020})\BibitemShut
  {NoStop}%
\bibitem [{Sup()}]{Supplemental}%
  \BibitemOpen
  \href@noop {} {}\bibinfo {note} {See Supplemental Material at [URL] for a
  derivation of the piezoelectric vector $\bm{g}(\bm{r})$ for substrates and
  interfaces, and a derivation of Eq.~(12). It also contains the material
  parameters used in Tables I and II which were extracted from experiments
  \cite{Fraga2014, Li1993, Kamm1964, Rubell1972, Hao2001}.}\BibitemShut {Stop}%
\bibitem [{\citenamefont {Wenner}\ \emph {et~al.}(2011)\citenamefont {Wenner},
  \citenamefont {Barends}, \citenamefont {Bialczak}, \citenamefont {Chen},
  \citenamefont {Kelly}, \citenamefont {Lucero}, \citenamefont {Mariantoni},
  \citenamefont {Megrant}, \citenamefont {O'Malley}, \citenamefont {Sank},
  \citenamefont {Vainsencher}, \citenamefont {Wang}, \citenamefont {White},
  \citenamefont {Yin}, \citenamefont {Zhao}, \citenamefont {Cleland},\ and\
  \citenamefont {Martinis}}]{Wenner2011}%
  \BibitemOpen
  \bibfield  {author} {\bibinfo {author} {\bibfnamefont {J.}~\bibnamefont
  {Wenner}}, \bibinfo {author} {\bibfnamefont {R.}~\bibnamefont {Barends}},
  \bibinfo {author} {\bibfnamefont {R.~C.}\ \bibnamefont {Bialczak}}, \bibinfo
  {author} {\bibfnamefont {Y.}~\bibnamefont {Chen}}, \bibinfo {author}
  {\bibfnamefont {J.}~\bibnamefont {Kelly}}, \bibinfo {author} {\bibfnamefont
  {E.}~\bibnamefont {Lucero}}, \bibinfo {author} {\bibfnamefont
  {M.}~\bibnamefont {Mariantoni}}, \bibinfo {author} {\bibfnamefont
  {A.}~\bibnamefont {Megrant}}, \bibinfo {author} {\bibfnamefont
  {P.}~\bibnamefont {O'Malley}}, \bibinfo {author} {\bibfnamefont
  {D.}~\bibnamefont {Sank}}, \bibinfo {author} {\bibfnamefont {A.}~\bibnamefont
  {Vainsencher}}, \bibinfo {author} {\bibfnamefont {H.}~\bibnamefont {Wang}},
  \bibinfo {author} {\bibfnamefont {T.}~\bibnamefont {White}}, \bibinfo
  {author} {\bibfnamefont {Y.}~\bibnamefont {Yin}}, \bibinfo {author}
  {\bibfnamefont {J.}~\bibnamefont {Zhao}}, \bibinfo {author} {\bibfnamefont
  {A.}~\bibnamefont {Cleland}}, \ and\ \bibinfo {author} {\bibfnamefont
  {J.}~\bibnamefont {Martinis}},\ }\href {\doibase 10.1063/1.3637047}
  {\bibfield  {journal} {\bibinfo  {journal} {Appl. Phys. Lett.}\ }\textbf
  {\bibinfo {volume} {99}},\ \bibinfo {pages} {113513} (\bibinfo {year}
  {2011})}\BibitemShut {NoStop}%
\bibitem [{\citenamefont {Georgescu}\ and\ \citenamefont
  {Ismail-Beigi}(2019)}]{Georgescu2019}%
  \BibitemOpen
  \bibfield  {author} {\bibinfo {author} {\bibfnamefont {A.~B.}\ \bibnamefont
  {Georgescu}}\ and\ \bibinfo {author} {\bibfnamefont {S.}~\bibnamefont
  {Ismail-Beigi}},\ }\href {\doibase 10.1103/PhysRevApplied.11.064065}
  {\bibfield  {journal} {\bibinfo  {journal} {Phys. Rev. Appl.}\ }\textbf
  {\bibinfo {volume} {11}},\ \bibinfo {pages} {064065} (\bibinfo {year}
  {2019})}\BibitemShut {NoStop}%
\bibitem [{\citenamefont {Tarumi}\ \emph {et~al.}(2007)\citenamefont {Tarumi},
  \citenamefont {Nakamura}, \citenamefont {Ogi},\ and\ \citenamefont
  {Hirao}}]{Tarumi2007}%
  \BibitemOpen
  \bibfield  {author} {\bibinfo {author} {\bibfnamefont {R.}~\bibnamefont
  {Tarumi}}, \bibinfo {author} {\bibfnamefont {K.}~\bibnamefont {Nakamura}},
  \bibinfo {author} {\bibfnamefont {H.}~\bibnamefont {Ogi}}, \ and\ \bibinfo
  {author} {\bibfnamefont {M.}~\bibnamefont {Hirao}},\ }\href {\doibase
  10.1063/1.2816252} {\bibfield  {journal} {\bibinfo  {journal} {J. Appl.
  Phys.}\ }\textbf {\bibinfo {volume} {102}},\ \bibinfo {pages} {113508}
  (\bibinfo {year} {2007})}\BibitemShut {NoStop}%
\bibitem [{\citenamefont {Paik}\ \emph {et~al.}(2011)\citenamefont {Paik},
  \citenamefont {Schuster}, \citenamefont {Bishop}, \citenamefont {Kirchmair},
  \citenamefont {Catelani}, \citenamefont {Sears}, \citenamefont {Johnson},
  \citenamefont {Reagor}, \citenamefont {Frunzio}, \citenamefont {Glazman},
  \citenamefont {Girvin}, \citenamefont {Devoret},\ and\ \citenamefont
  {Schoelkopf}}]{Paik2011}%
  \BibitemOpen
  \bibfield  {author} {\bibinfo {author} {\bibfnamefont {H.}~\bibnamefont
  {Paik}}, \bibinfo {author} {\bibfnamefont {D.~I.}\ \bibnamefont {Schuster}},
  \bibinfo {author} {\bibfnamefont {L.~S.}\ \bibnamefont {Bishop}}, \bibinfo
  {author} {\bibfnamefont {G.}~\bibnamefont {Kirchmair}}, \bibinfo {author}
  {\bibfnamefont {G.}~\bibnamefont {Catelani}}, \bibinfo {author}
  {\bibfnamefont {a.~P.}\ \bibnamefont {Sears}}, \bibinfo {author}
  {\bibfnamefont {B.~R.}\ \bibnamefont {Johnson}}, \bibinfo {author}
  {\bibfnamefont {M.~J.}\ \bibnamefont {Reagor}}, \bibinfo {author}
  {\bibfnamefont {L.}~\bibnamefont {Frunzio}}, \bibinfo {author} {\bibfnamefont
  {L.~I.}\ \bibnamefont {Glazman}}, \bibinfo {author} {\bibfnamefont {S.~M.}\
  \bibnamefont {Girvin}}, \bibinfo {author} {\bibfnamefont {M.~H.}\
  \bibnamefont {Devoret}}, \ and\ \bibinfo {author} {\bibfnamefont {R.~J.}\
  \bibnamefont {Schoelkopf}},\ }\href {\doibase 10.1103/PhysRevLett.107.240501}
  {\bibfield  {journal} {\bibinfo  {journal} {Phys. Rev. Lett.}\ }\textbf
  {\bibinfo {volume} {107}},\ \bibinfo {pages} {240501} (\bibinfo {year}
  {2011})}\BibitemShut {NoStop}%
\bibitem [{\citenamefont {Lin}\ \emph {et~al.}(2018)\citenamefont {Lin},
  \citenamefont {Nguyen}, \citenamefont {Grabon}, \citenamefont {{San Miguel}},
  \citenamefont {Pankratova},\ and\ \citenamefont {Manucharyan}}]{Lin2018}%
  \BibitemOpen
  \bibfield  {author} {\bibinfo {author} {\bibfnamefont {Y.~H.}\ \bibnamefont
  {Lin}}, \bibinfo {author} {\bibfnamefont {L.~B.}\ \bibnamefont {Nguyen}},
  \bibinfo {author} {\bibfnamefont {N.}~\bibnamefont {Grabon}}, \bibinfo
  {author} {\bibfnamefont {J.}~\bibnamefont {{San Miguel}}}, \bibinfo {author}
  {\bibfnamefont {N.}~\bibnamefont {Pankratova}}, \ and\ \bibinfo {author}
  {\bibfnamefont {V.~E.}\ \bibnamefont {Manucharyan}},\ }\href {\doibase
  10.1103/PhysRevLett.120.150503} {\bibfield  {journal} {\bibinfo  {journal}
  {Phys. Rev. Lett.}\ }\textbf {\bibinfo {volume} {120}},\ \bibinfo {pages}
  {150503} (\bibinfo {year} {2018})}\BibitemShut {NoStop}%
\bibitem [{\citenamefont {Earnest}\ \emph {et~al.}(2018)\citenamefont
  {Earnest}, \citenamefont {Chakram}, \citenamefont {Lu}, \citenamefont
  {Irons}, \citenamefont {Naik}, \citenamefont {Leung}, \citenamefont {Ocola},
  \citenamefont {Czaplewski}, \citenamefont {Baker}, \citenamefont {Lawrence},
  \citenamefont {Koch},\ and\ \citenamefont {Schuster}}]{Earnest2018}%
  \BibitemOpen
  \bibfield  {author} {\bibinfo {author} {\bibfnamefont {N.}~\bibnamefont
  {Earnest}}, \bibinfo {author} {\bibfnamefont {S.}~\bibnamefont {Chakram}},
  \bibinfo {author} {\bibfnamefont {Y.}~\bibnamefont {Lu}}, \bibinfo {author}
  {\bibfnamefont {N.}~\bibnamefont {Irons}}, \bibinfo {author} {\bibfnamefont
  {R.~K.}\ \bibnamefont {Naik}}, \bibinfo {author} {\bibfnamefont
  {N.}~\bibnamefont {Leung}}, \bibinfo {author} {\bibfnamefont
  {L.}~\bibnamefont {Ocola}}, \bibinfo {author} {\bibfnamefont {D.~A.}\
  \bibnamefont {Czaplewski}}, \bibinfo {author} {\bibfnamefont
  {B.}~\bibnamefont {Baker}}, \bibinfo {author} {\bibfnamefont
  {J.}~\bibnamefont {Lawrence}}, \bibinfo {author} {\bibfnamefont
  {J.}~\bibnamefont {Koch}}, \ and\ \bibinfo {author} {\bibfnamefont {D.~I.}\
  \bibnamefont {Schuster}},\ }\href {\doibase 10.1103/PhysRevLett.120.150504}
  {\bibfield  {journal} {\bibinfo  {journal} {Phys. Rev. Lett.}\ }\textbf
  {\bibinfo {volume} {120}},\ \bibinfo {pages} {150504} (\bibinfo {year}
  {2018})}\BibitemShut {NoStop}%
\bibitem [{\citenamefont {Fraga}\ \emph {et~al.}(2014)\citenamefont {Fraga},
  \citenamefont {Furlan}, \citenamefont {Pessoa},\ and\ \citenamefont
  {Massi}}]{Fraga2014}%
  \BibitemOpen
  \bibfield  {author} {\bibinfo {author} {\bibfnamefont {M.~A.}\ \bibnamefont
  {Fraga}}, \bibinfo {author} {\bibfnamefont {H.}~\bibnamefont {Furlan}},
  \bibinfo {author} {\bibfnamefont {R.}~\bibnamefont {Pessoa}}, \ and\ \bibinfo
  {author} {\bibfnamefont {M.}~\bibnamefont {Massi}},\ }\href {\doibase
  10.1007/s00542-013-2029-z} {\bibfield  {journal} {\bibinfo  {journal}
  {Microsyst. Technol.}\ }\textbf {\bibinfo {volume} {20}},\ \bibinfo {pages}
  {9} (\bibinfo {year} {2014})}\BibitemShut {NoStop}%
\bibitem [{\citenamefont {Li}\ \emph {et~al.}(1993)\citenamefont {Li},
  \citenamefont {Grimsditch}, \citenamefont {Xu},\ and\ \citenamefont
  {Chan}}]{Li1993}%
  \BibitemOpen
  \bibfield  {author} {\bibinfo {author} {\bibfnamefont {Z.}~\bibnamefont
  {Li}}, \bibinfo {author} {\bibfnamefont {M.}~\bibnamefont {Grimsditch}},
  \bibinfo {author} {\bibfnamefont {X.}~\bibnamefont {Xu}}, \ and\ \bibinfo
  {author} {\bibfnamefont {S.~K.}\ \bibnamefont {Chan}},\ }\href {\doibase
  10.1080/00150199308223459} {\bibfield  {journal} {\bibinfo  {journal}
  {Ferroelectrics}\ }\textbf {\bibinfo {volume} {141}},\ \bibinfo {pages} {313}
  (\bibinfo {year} {1993})}\BibitemShut {NoStop}%
\bibitem [{\citenamefont {Kamm}\ and\ \citenamefont {Alers}(1964)}]{Kamm1964}%
  \BibitemOpen
  \bibfield  {author} {\bibinfo {author} {\bibfnamefont {G.~N.}\ \bibnamefont
  {Kamm}}\ and\ \bibinfo {author} {\bibfnamefont {G.~A.}\ \bibnamefont
  {Alers}},\ }\href {\doibase 10.1063/1.1713309} {\bibfield  {journal}
  {\bibinfo  {journal} {J. Appl. Phys.}\ }\textbf {\bibinfo {volume} {35}},\
  \bibinfo {pages} {327} (\bibinfo {year} {1964})}\BibitemShut {NoStop}%
\bibitem [{\citenamefont {Rubell}\ and\ \citenamefont
  {Brotzen}(1972)}]{Rubell1972}%
  \BibitemOpen
  \bibfield  {author} {\bibinfo {author} {\bibfnamefont {W.~C.}\ \bibnamefont
  {Rubell}}\ and\ \bibinfo {author} {\bibfnamefont {F.~R.}\ \bibnamefont
  {Brotzen}},\ }\href {\doibase 10.1063/1.1661712} {\bibfield  {journal}
  {\bibinfo  {journal} {J. Appl. Phys.}\ }\textbf {\bibinfo {volume} {43}},\
  \bibinfo {pages} {3306} (\bibinfo {year} {1972})}\BibitemShut {NoStop}%
\bibitem [{\citenamefont {Hao}\ and\ \citenamefont {Maris}(2001)}]{Hao2001}%
  \BibitemOpen
  \bibfield  {author} {\bibinfo {author} {\bibfnamefont {H.~Y.}\ \bibnamefont
  {Hao}}\ and\ \bibinfo {author} {\bibfnamefont {H.~J.}\ \bibnamefont
  {Maris}},\ }\href {\doibase 10.1103/PhysRevB.63.224301} {\bibfield  {journal}
  {\bibinfo  {journal} {Phys. Rev. B}\ }\textbf {\bibinfo {volume} {63}},\
  \bibinfo {pages} {224301} (\bibinfo {year} {2001})}\BibitemShut {NoStop}%
\end{thebibliography}%

\end{document}